\documentclass[aps,prl,twocolumn,showpacs,superscriptaddress]{revtex4}
\usepackage{graphicx}

\begin{document}

\title{Emergence of Scale-Free Networks from Local Connectivity and
Communication Trade-offs}

\author{Valmir C. Barbosa}
\affiliation{Programa de Engenharia de Sistemas e Computa\c c\~ao, COPPE,
Universidade Federal do Rio de Janeiro,
Caixa Postal 68511, 21941-972 Rio de Janeiro - RJ, Brazil}

\author{Raul Donangelo}
\author{Sergio R. Souza}
\affiliation{Instituto de F\'\i sica,
Universidade Federal do Rio de Janeiro,
Caixa Postal 68528, 21941-972 Rio de Janeiro - RJ, Brazil}

\date{\today}

\begin{abstract}
We introduce a new mechanism of connectivity evolution in networks to account
for the emergence of scale-free behavior. The mechanism works on a fixed set of
nodes and promotes growth from a minimally connected initial topology by the
addition of edges. A new edge is added between two nodes depending on the
trade-off between a gain and a cost function of local connectivity and
communication properties. We report on simulation results that indicate the
appearance of power-law distributions of node degrees for selected parameter
combinations.
\end{abstract}


\maketitle

The topology of large-scale complex networks such as the Internet and the WWW is
in general not known. The study of such networks has then relied on modeling
them as random graphs \cite{b01}, and particularly on focusing almost
exclusively on the distribution of node degrees. Unlike the classic case
pioneered by Erd\H{o}s and R\'{e}nyi \cite{er59}, for the Internet, the WWW, and
several other networks, it appears that degrees are distributed according to a
power law, not to a Poisson distribution. That is, the probability that a
randomly chosen node has degree $k$ is proportional to $k^{-\tau}$, in general
with $2<\tau<3$ \cite{ab02,dm02}.

These findings are based on probe samplers in the case of both the Internet
\cite{fff99} and the WWW \cite{baj00}, and are generally regarded as reasonably
accurate. However, part of the underlying machinery has been recently proven
somewhat unreliable in the case of the Internet. For example, it has been
demonstrated experimentally that the usual mechanism of inferring
breadth-first-search trees from the probe results can underestimate the value of
$\tau$ significantly when the graph does have a power-law degree distribution
\cite{cm05}. Likewise, it is possible to argue formally that such a mechanism
can in some cases lead to the conclusion of a power-law degree distribution when
in fact the graph's degrees are distributed in some other way \cite{ackm05}.

In recent years, and notwithstanding these limitations, considerable effort has
been put into discovering mechanisms of network growth that give rise to a
power-law degree distribution. Especially noteworthy is the mechanism of
preferential attachment, which underlies the so-called Barab\'{a}si-Albert model
\cite{ba99,baj99}, as well as variations \cite{dms00,krl00,hk02,br04} and
generalizations \cite{cf03} thereof. Preferential attachment is the policy
whereby a new edge is added to the network between a new node and a pre-existing
one with probability proportional to how many edges are already incident on the
pre-existing node, that is, its current degree. The generalization of
\cite{cf03} incorporates both this policy and also the copying mechanism of
\cite{krrstu00}. We refer the reader to \cite{br03} for a review of the
essential mathematical results related to these models.

While the study of complex networks from the perspective of node-degree
distributions seems sound and has given rise to important discoveries related to
global properties, such as the nature and size of a network's connected
components and diameter \cite{nsw01}, explaining the formation of the network
from the same perspective (e.g., by evoking preferential attachment) is
unreasonable for at least two reasons. The first is that the addition of a
particular edge most definitely does not depend on global properties such as the
distribution of node degrees at the time of expansion. The second reason is
that, at least for computer networks like the Internet, it makes no sense to
assume that the degree distribution, rather than some cost- or
performance-related entity, is the essential driving force behind the evolution
of the network's topology.

Models that depend on node-degree distributions are then adequate descriptive
models, in the sense that they give rise to the desired power-law functional
form, but constitute poor generative models. This has also been recognized
elsewhere (cf., e.g., \cite{tgjsw02,lawd04}), and has resulted in the appearance
of alternative models, such as the ones in \cite{fkp02,flmps03}. These, however,
are also dependent on global properties, such as one-to-all distances, and
therefore seem implausible as well.

We work on the premise that networks such as the Internet or the WWW, although
fast-growing, appear not to acquire new nodes fast enough to impact their main
topological properties significantly. Thus the model that we study is targeted
at the evolution of the connectivity of computer networks, and promotes network
growth on a fixed set of nodes by incrementally adding edges between nodes as
the result of comparing a gain function and a cost function for each edge
addition. If $i$ and $j$ are nodes not currently connected by an edge in the
network, the gain incurred when adding an edge between them depends only on the
immediate neighborhoods of $i$ and $j$ and on the current distance between $i$
and $j$. The cost of the addition, in turn, is also dependent solely upon $i$
and $j$ and seeks to reflect both the cost of deploying the communications link
itself and the cost of upgrading nodes $i$ and $j$'s connection capabilities to
accommodate the new link. The edge joining $i$ and $j$ is added to the network
if the gain surpasses the cost.

We model the evolution of network connectivity as the sequence
$G^0,G^1,\ldots$ of undirected graphs, all having the same set of $n$ nodes. We
assume that $G^0$ is a tree that spans all the nodes; $G^0$ is therefore
connected and has $n-1$ edges. For $t\ge 0$, $G^{t+1}$ is obtained from $G^t$ by
randomly selecting two nodes, say $i$ and $j$, that are not directly connected
by an edge, and then adding an edge between them if the gain incurred with the
addition of the edge is greater than its cost. Otherwise, we simply let
$G^{t+1}=G^t$. All graphs in the sequence are then guaranteed to be connected
and to remain free of multiple edges and self-loops. We let $d^t_{ij}$ denote
the distance between $i$ and $j$ in $G^t$, and $n^t_i$ the degree of node $i$ in
$G^t$. We also let $N^t_i(j)$ be the set comprising every neighbor $k$ of node
$i$ in $G^t$ for which $d^t_{jk}>2$, and similarly $N^t_{ij}$ be the set of
unordered node pairs $(k,l)$ such that either $k$ or $l$ is a neighbor of $i$,
the other node in the pair is a neighbor of $j$, and furthermore $d^t_{kl}>3$.

Let $g^t_{ij}$ denote the gain incurred with the addition of an edge between $i$
and $j$ to $G^t$ when $d^t_{ij}>1$. In our model, we let $g^t_{ij}$ be some
upper bound on the number of edges by which distances between certain nodes
become shorter after the addition of that edge. The distances we consider to
establish this upper bound are some of those that involve $i$ or $j$ directly,
or yet nodes in their immediate neighborhoods in $G^t$. Specifically, we
consider $d^t_{ij}$, $d^t_{ik}$ for $k\in N^t_j(i)$ (neighbors of $j$ that are
more than $2$ edges away from $i$ in $G^t$), $d^t_{jk}$ for $k\in N^t_i(j)$
(neighbors of $i$ that are more than $2$ edges away from $j$ in $G^t$), and
finally $d^t_{kl}$ for $(k,l)\in N^t_{ij}$ (node pairs that are more than $3$
edges away from each other in $G^t$, one being a neighbor of $i$, the other a
neighbor of $j$).

Upper bounds on each distance in the latter three groups are, clearly,
$d^t_{ij}+1$, $d^t_{ij}+1$, and $d^t_{ij}+2$, respectively. An upper bound on
the sum of all distances considered is then
\begin{equation}
d^t_{ij}+
(d^t_{ij}+1)\vert N^t_j(i)\vert+
(d^t_{ij}+1)\vert N^t_i(j)\vert+
(d^t_{ij}+2)\vert N^t_{ij}\vert,
\end{equation}
where we use $\vert X\vert$ to denote the cardinality of set $X$. The addition
of an edge joining $i$ and $j$ causes the sum of all these distances to become
\begin{equation}
1+
2\vert N^t_j(i)\vert+
2\vert N^t_i(j)\vert+
3\vert N^t_{ij}\vert,
\end{equation}
and consequently the overall number of edges by which the distances become
shorter is at most
\begin{equation}
\label{gain}
g^t_{ij}=(d^t_{ij}-1)
\left[1+
\vert N^t_j(i)\vert+
\vert N^t_i(j)\vert+
\vert N^t_{ij}\vert
\right].
\end{equation}

One crucial aspect of the gain expressed in (\ref{gain}) is that, in the
context of computer networks, it depends exclusively on information that can be
obtained by tracing routes on $G^t$. This certainly holds for the determination
of $d^t_{ij}$, and holds also for determining the sets $N^t_j(i)$, $N^t_i(j)$,
and $ N^t_{ij}$, provided only that the process of tracing routes is controlled
for constant depth. However, given the nature of routing algorithms such as
those of the Internet \cite{kr05}, tracing a route between $i$ and $j$ on $G^t$
is only guaranteed to provide an upper bound on $d^t_{ij}$, which is nonetheless
consonant with the expression in (\ref{gain}) being itself an upper bound on
total distance improvement.

Besides $g^t_{ij}$, the decision regarding the addition of an edge between $i$
and $j$ depends also on the cost of this addition. We denote this cost by
$c^t_{ij}$ and define it in such a way that both the cost of deploying a
communications link and the cost of possibly upgrading the connection
capabilities of $i$ or $j$ are taken into account. The former of these we denote
by $C$ and assume to be independent of $t$, $i$, or $j$.

As for the latter of the two cost components, we assume that the number of
connections a node can sustain at any time is at most $\lceil\alpha^z\rceil$ for
some fixed $\alpha>1$ and some $z\in\{0,1,\ldots\}$ that does not decrease as
time elapses (we use $\lceil x\rceil$ to denote the least integer that is no
less than $x$). If the degree of $i$ or $j$ in $G^t$ is precisely such a maximum
number of connections, then the cost of connecting $i$ to $j$ directly involves
the cost of upgrading the connection capabilities of $i$ or $j$, as the case may
be, to $\lceil\alpha^{z+w}\rceil$, where $w$ is the least integer for which
$\lceil\alpha^{z+w}\rceil>\lceil\alpha^z\rceil$. We further assume, for some
fixed $\beta>1$, that the cost of endowing the node with the capability of
connecting to $\lceil\alpha^z\rceil$ other nodes is proportional to $\beta^z$.

Let $n_k$ denote the number of connections of some node $k$. We model the
scenario in which the cost incurred with the upgrade of $n_k$ from
$\lceil\alpha^z\rceil$ to $\lceil\alpha^{z+w}\rceil$ is proportional to
$\beta^{z+w}-\beta^z$ (only the cost difference is paid) and is furthermore
amortized along the deployment of each new connection (as opposed to being paid
in full when the $\lceil\alpha^z\rceil+1$st connection is deployed). If we let
$f(n_k)$ be the cost portion to be incurred when the number of connections is
$n_k$ and for simplicity disregard the fact that connections necessarily occur
in discrete numbers, then it follows that
\begin{eqnarray}
\int_{\alpha^z}^{\alpha^{z+w}}f(n_k)\,dn_k
&\propto&\beta^{z+w}-\beta^z\\
&=&(\alpha^{z+w})^{\log_\alpha\beta}-(\alpha^z)^{\log_\alpha\beta}.
\end{eqnarray}
Consequently,
\begin{equation}
f(n_k)\propto n_k^{\log_\alpha(\beta/\alpha)}.
\end{equation}
Setting $\alpha=\beta$ leads $f(n_k)$ to be constant with respect to $n_k$;
setting $\alpha\neq\beta$ leads $f(n_k)$ to vary either directly
($\alpha<\beta$) or inversely ($\alpha>\beta$) with $n_k$.

We then have
\begin{equation}
\label{cost}
c^t_{ij}=C+D\left[(n^t_i)^\gamma+(n^t_j)^\gamma\right]
\end{equation}
for some constant $D$ and $\gamma=\log_\alpha(\beta/\alpha)$. An edge is added
to $G^t$ between nodes $i$ and $j$ to yield $G^{t+1}$ if $g^t_{ij}>c^t_{ij}$. If
not, then $G^{t+1}=G^t$. By the nature of (\ref{gain}) and (\ref{cost}), this
decision involves only the distance between $i$ and $j$ in $G^t$, in addition to
other quantities that depend exclusively on the surroundings of $i$ and $j$
within a constant radius in $G^t$. It is then essentially a local decision.

We have conducted computer simulations for selected combinations of the $C$,
$D$, and $\gamma$ parameters. Each simulation starts with a randomly chosen
instance of $G^0$ and proceeds through $t=3000n$. A $G^0$ instance is generated
on the $n$ initially isolated nodes by progressively selecting node pairs at
random and directly interconnecting them if no path exists between them; because
$G^0$ is a tree, it is necessary and sufficient that $n-1$ such interconnections
be performed.

\begin{figure}
\vspace{0.55cm}
\includegraphics[scale=0.38]{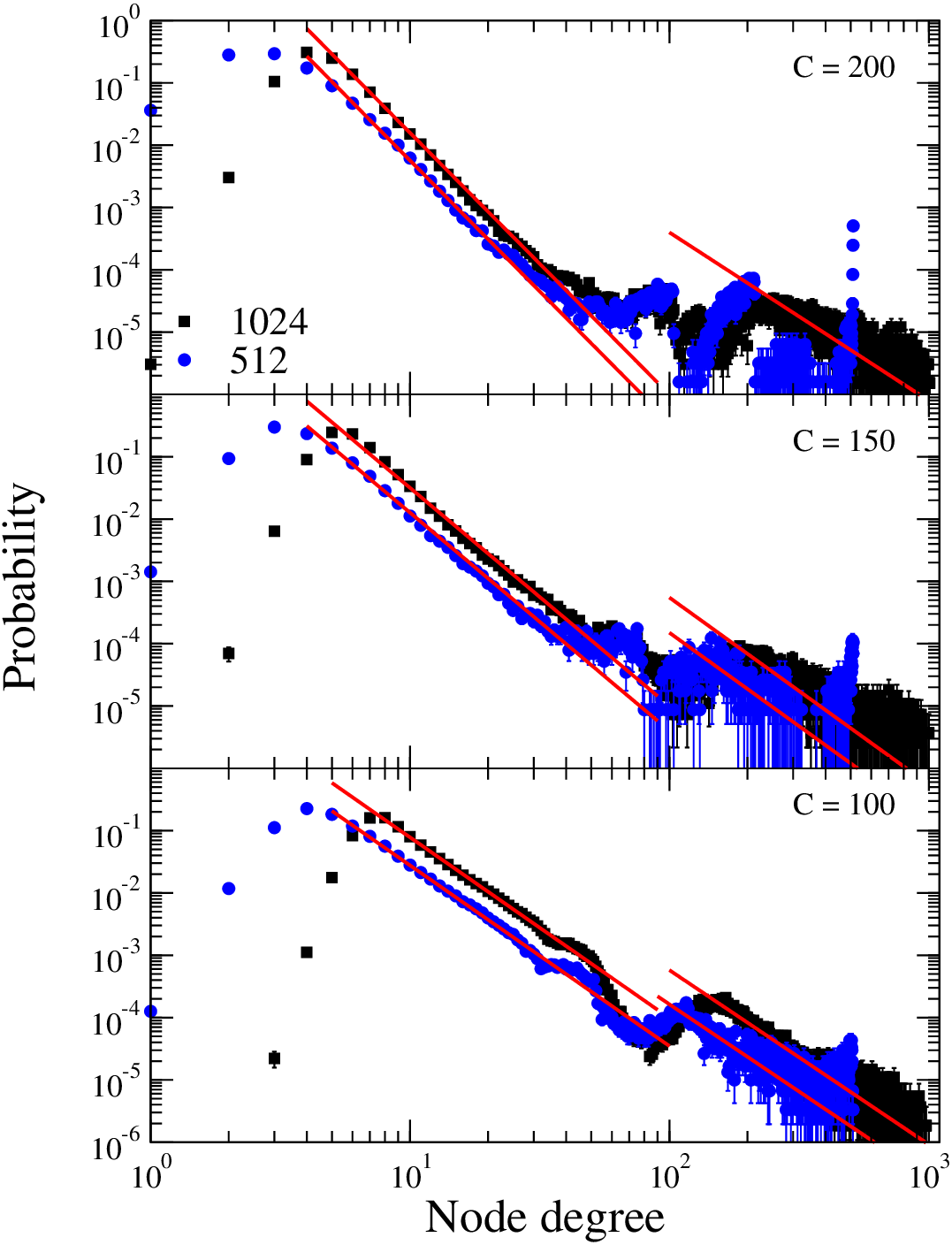}
\caption{(Color online) Average node-degree distributions for $n=512,1024$,
$D=0.1$, and $\gamma=0.9$. Values for $\tau$ are between $2.9$ and $4.2$ for the
lower degrees, $2.7$ and $3.0$ for the higher degrees.}
\label{figC}
\end{figure}

\begin{figure}
\vspace{0.55cm}
\includegraphics[scale=0.38]{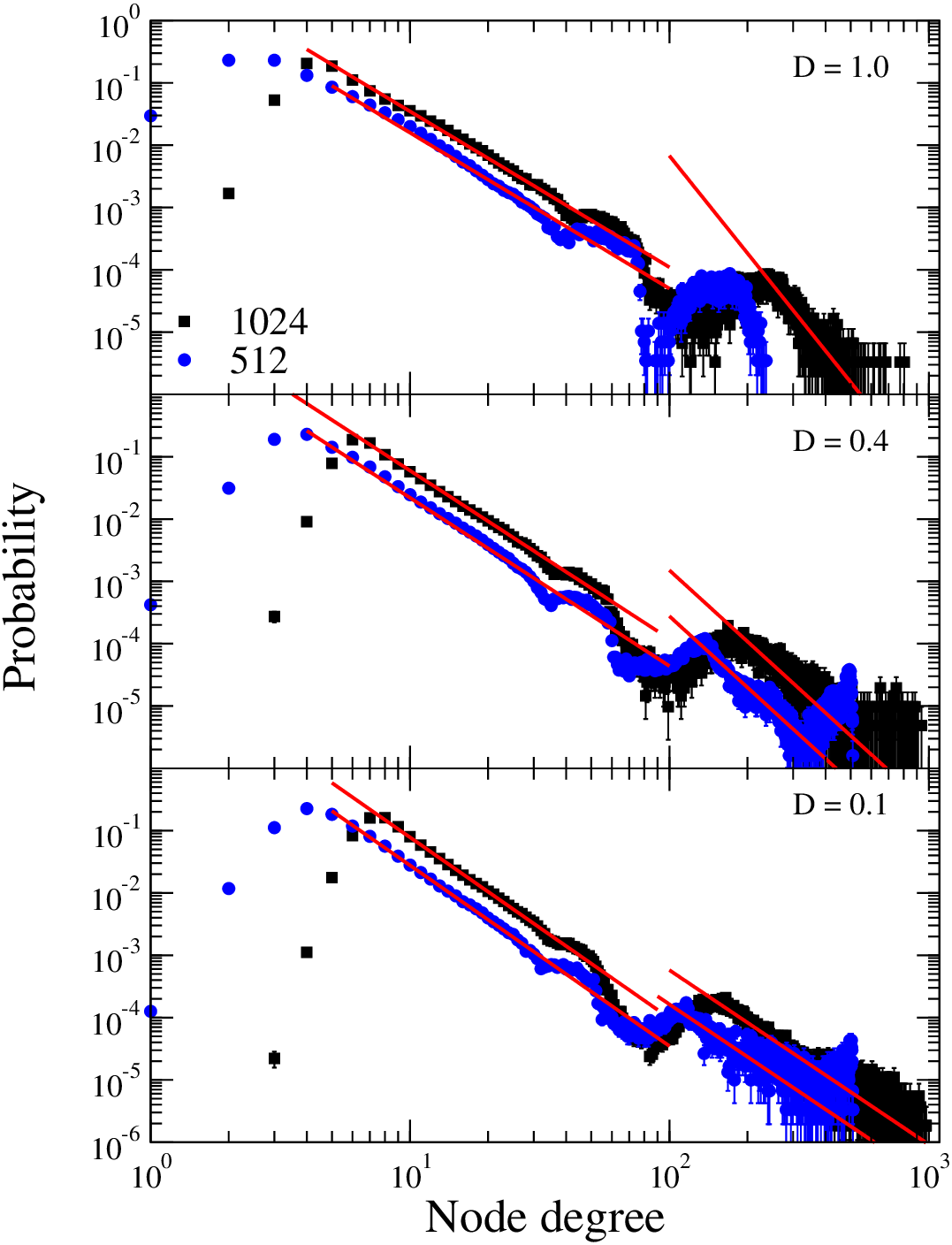}
\caption{(Color online) Average node-degree distributions for $n=512,1024$,
$C=100$, and $\gamma=0.9$. Values for $\tau$ are between $2.5$ and $2.9$ for the
lower degrees, $2.9$ and $5.2$ for the higher degrees.}
\label{figD}
\end{figure}

At each step of a simulation the distance $d^t_{ij}$ must be calculated on
$G^t$. While on a real computer network such a distance (or an upper bound
thereof) is readily available from the network's routing structure (as noted
earlier), calculating $d^t_{ij}$ seems to be asymptotically no easier than
finding the distances between a given node and all others in $G^t$. For
connected graphs, this requires $O(m^t)$ time \cite{clrs01}, where we use $m^t$
to denote the number of edges of $G^t$. It is therefore a time-consuming
procedure, and progressively more so as the simulation is carried on and the
graph tends to become denser. The consequence of this for the present study is
that the value of $n$ is somewhat limited, and so is the number of independent
$G^0$ instances that can be used for statistical significance.

\begin{figure}
\includegraphics[scale=0.38]{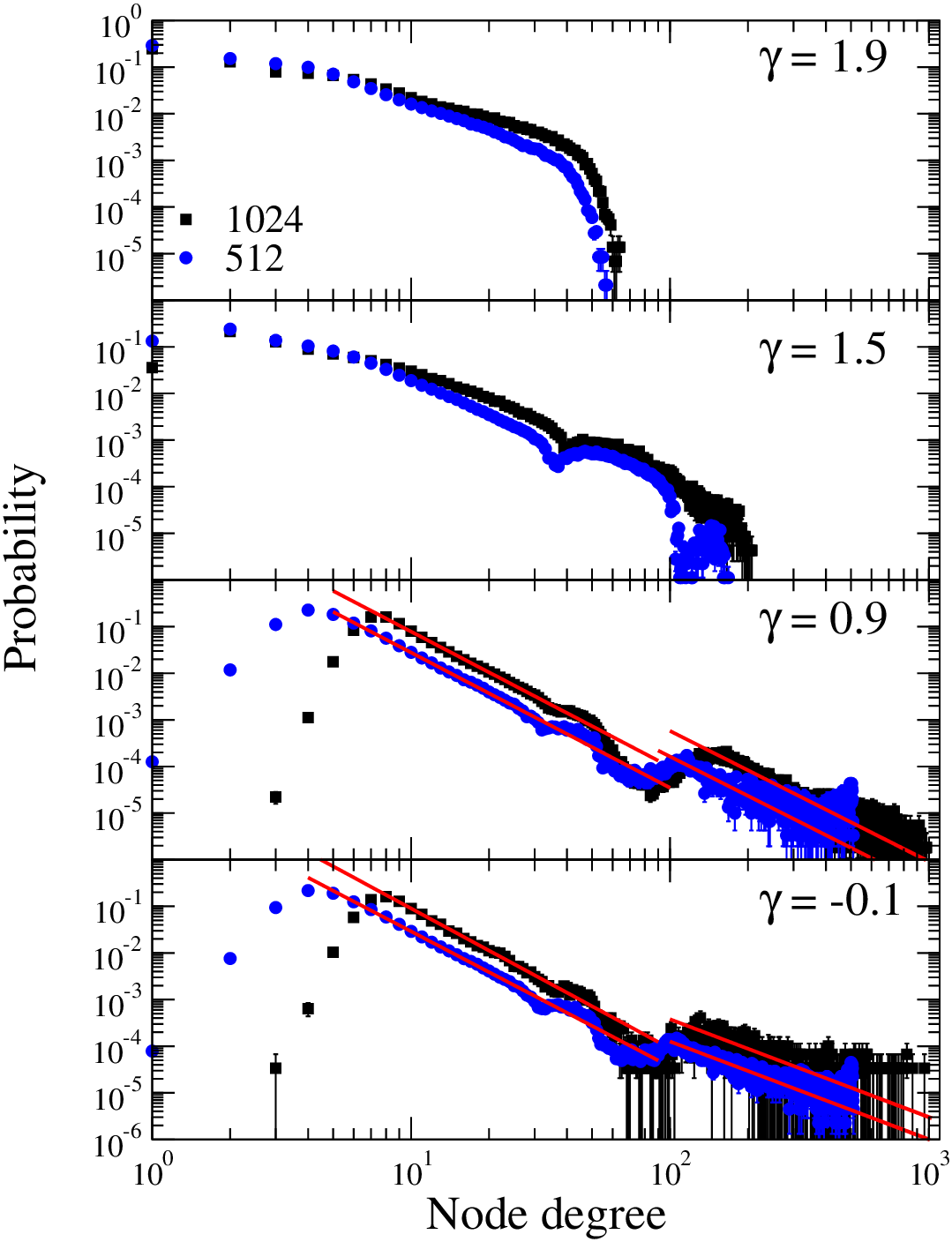}
\caption{(Color online) Average node-degree distributions for $n=512,1024$,
$C=100$, and $D=0.1$. Values for $\tau$ are between $2.9$ and $3.0$ for the
lower degrees, $2.1$ and $2.9$ for the higher degrees.}
\label{figgamma}
\end{figure}

Our results are shown in Figures~\ref{figC}--\ref{figgamma}, where,
respectively, the value of each of $C$, $D$, and $\gamma$ is varied while the
other two parameters remain fixed at a set of common values ($C=100$, $D=0.1$,
and $\gamma=0.9$). For each combination of the three parameters we show results
for two values of $n$. As the figures indicate, our model for network growth
does indeed give rise to a scale-free pattern of behavior in which a vast
majority of the nodes has low degrees while a few high-degree nodes are
nonetheless present.

The figures also indicate, except for the highest $\gamma$ values in our
simulations ($\gamma=1.5,1.9$, cf.\ Figure~\ref{figgamma}), that the node-degree
distribution seems to settle at two distinct power-law regimes, one for node
degrees below roughly $100$, the other for those above this threshold. While a
definitive explanation of why this happens depends upon a more detailed analysis
of the process whereby nodes acquire ever higher degrees, we conjecture that it
may go along the following lines.

In our model, nodes acquire higher degrees one unit at a time when two of them
become directly connected to each other as a result of comparing the gain in
(\ref{gain}) to the cost in (\ref{cost}). As degrees become larger and the
network denser, it also happens that distances between node pairs become
shorter. The node sets whose cardinalities appear in (\ref{gain}) tend,
therefore, to become smaller. Together, these trends make it progressively
harder for gains to surpass costs and for degrees to continue increasing.

However, a few high-degree nodes do appear and the dynamics of network growth
may occasionally consider joining two of them together. Because they have high
degrees, it may happen that the node-set cardinalities in (\ref{gain}) become
once again relatively non-negligible and a few high-degree node pairs do indeed
become interconnected. Our results indicate that, if this is what happens, then
its occurrence inaugurates a new power-law regime for the highest degrees. In
this case, what we witness may be the emergence of some sort of hierarchical
organization within the network, not unlike what happens with the Internet
\cite{kr05}, which is inherently organized in just such a way (that a single
power-law regime should be reported in topology measurements like those of
\cite{fff99} may be due exclusively to the fact that they are constrained to
within one single level of the hierarchy).

\begin{acknowledgments}
We acknowledge partial support from CNPq, CAPES, FAPERJ BBP grants, and the
PRONEX initiative under contract PRONEX/FAJER 26.171.176.2003.
\end{acknowledgments}

\bibliography{plnet}
\bibliographystyle{apsrev}

\end{document}